\documentclass[12pt,preprint,longabstract]{aastex}

\usepackage{graphicx}
\usepackage{txfonts}
\usepackage{natbib}

\newcommand{\kpc}{~h^{-1}~kpc}
\newcommand{\Mpc}{~h^{-1}~Mpc}
\newcommand{\Msun}{~h^{-1}~M_{\odot}}
\newcommand{\den}{~h^{3}~Mpc^{-3}}

\newcommand{\gtsim}{~ _{\sim}^{>} ~}
\slugcomment{}

\shorttitle{Galactic Haloes and their Evolution}

\shortauthors{ Jian \& Chiueh \& Chien}

\begin{document}


\title{Galactic Halos Derived from $\Lambda CDM$ Cosmology Simulation and their Red-Shift Evolution}

\author{Hung-Yu \textsc{Jian, Chia-Hung \textsc{Chien}} %
  }
\affil{Department of Physics, National Taiwan University, 106, Taipei, Taiwan, R.O.C.}
\email{hyj@phys.ntu.edu.tw}
\and
\author{Tzihong \textsc{Chiueh}}
\affil{Department of Physics, National Taiwan University, 106, Taipei, Taiwan, R.O.C. \\
       Center for Theoretical Sciences, National Taiwan University, 106, Taipei, Taiwan, R.O.C. \\
       LeCosPa, National Taiwan University, 106, Taipei, Taiwan, R.O.C. }
\email{chiuehth@phys.ntu.edu.tw}

%




%

\begin{abstract}
   Galaxies can form in a sufficiently deep gravitational potential so that
efficient gas cooling occurs. We estimate that such potential is provided
by a halo of mass $M \gtsim M_{c} \approx 7.0 \times 10^{12} ~
(\Delta_{c}(z) (1+z)^{3})^{-1/2} \Msun$, where $\Delta_{c}(z)$ is
the mean overdensity of spherically virialized objects formed at
redshift $z$, and $M_{c} \approx 4.0 \times 10^{11} \Msun$ at $z =
0$. Based on this criterion, our galaxy samples are constructed from cosmology simulation data by using HiFOF to select subhalos in those FOF halos that are more massive than $M_{c}$. There are far more dark subhalos than galaxy-hosting subhalos. Several tests against observations have been performed to examine our galaxy samples, including the differential galaxy mass functions, the galaxy space density, the projected
two point correlation functions (CF), the HODs, and the kinematic
pair fractions. These tests show good agreements.
Based on the consistency with observations, our galaxy
sample is believed to correctly represent galaxies in real universe,
and can be used to study other unexplored galaxy properties.

\end{abstract}

   {}
   {}
   {}
   {}


%

\section{Introduction}

The standard theoretical model for galaxy formation basically
consists of two main elements, the cold dark matter (CDM) model and
a dark energy field (which may take the form of a cosmological
constant, $\Lambda$). The fundamental assumption made in the theory
is that structure grew from weak density fluctuations present in the
otherwise homogeneous and rapidly expanding early Universe.
Afterwards, the gravitational instability drives these fluctuations into nonlinear regime in a bottom up fashion and they gradually grow to a wealth of structures today. Recent diverse cosmological studies, no matter from
large-scale structure observations \citep[e.g., ][]{spe03}, from
supernova data \citep[e.g., ][]{rie98}, or from light element
abundance \citep[e.g., ][]{whi93}, all seem to support the
concordance model. Due to the highly nonlinear nature in the
collapse of fluctuations and the subsequent hierarchical build-up of
structure, numerical simulations come to play an indispensable role.

In the past thirty years, galaxy clustering has been intensively
studied, as the clustering of galaxies has long been an essential
testing ground for various cosmological models and galaxy formation
scenarios. Especially due to the advent of modern computers, high
resolution simulations become possible. Cosmological N-body
simulations have developed into a powerful tool for calculating the
gravitational clustering of collisionless dark matter from specified
initial conditions. As the resolution of simulations increases, some
numerical problems such as the overmerging problem \citep{kly99} can
be overcome, and the galactic size scale in a large scale structure
simulation can be resolvable.
 In addition, the growth of galaxy redshift surveys has led to measurements of
increasing precision and detail. All these works have made the
comparisons between theories and observations possible.

There are four main approaches used in cosmological simulations to
study the galaxy clustering and properties. The first one is N-body
plus hydrodynamical simulation including gas and dark matter
particles \citep[e.g., ][]{wei04}. The second method is a hybrid
method that combines N-body simulations of the dark matter component
with semi-analytic treatments of the galaxy formation physics
\citep[e.g., ][]{spr05}. Another approach is using high-resolution
collisionless N-body simulations that identify galaxies with
"subhalos" in the dark matter distribution \citep[e.g.,
][]{kly99,kra04,con06}. The last one is the so called Halo
Occupation Distribution (HOD) approach which gives a purely
statistical description of how dark matter halos are populated with
galaxies \citep[e.g., ][]{ber03,kra04,zhe05}. It is known that the
cold dark matter is the dominant mass component and it interacts
only through gravity. To some extent, gravitational dynamics alone
should explain the basic features of galaxy clustering. \cite{ney04}
tried to understand how the infrared-selected galaxies populate dark
matter halos, paying special attention to the method of halo
identification in simulations. They tested the hypothesis that
baryonic physics negligibly affects the distribution of subhalos
down to the smallest scales yet observed and successfully reproduced
the Point Source Catalogue Redshift (PSCz) power spectrum.
\cite{ber03} reported that the HOD prediction of two methods, a
semi-analytic model and gasdynamics simulations, agree remarkably
well for samples of the same space density. This result indirectly
supported the idea that the HOD, and hence galaxy clustering, is
driven primarily by gravitational dynamics rather than by processes
such as cooling and star formation. On the other hand, \cite{kra04}
adopt a variant of the bound density maxima (BDM) halo-finding
algorithm \citep{kly99} to identify halos and subhalos and use the
maximum circular velocity $V_{max}$ as a proxy of halo mass for
selecting galaxy sample. Instead of selecting objects in a given
range of $V_{max}$, at each epoch they selected objects of a set of
number densities consistent with observational number densities and
corresponding to (redshift dependent) thresholds in maximum circular
velocity, i.e., $n_{i}(>V_{max})$. Their results show that the
dependence of correlation amplitude on the galaxy number density in
their sample is in general agreement with results from the Sloan
Digital Sky Survey. \cite{con06} instead use the maximum circular
velocity at the time of accretion, $V^{acc}_{max}$, for subhalos,
and the results show agreement with the observed galaxy clustering
in the SDSS data at $z \approx 0$ and in the DEEP2 samples at $z
\approx 1$ over the range of separations, $0.1 < r_{p}/(h^{-1} Mpc)
< 10.0$. However, this work lacks the information of the subhalo
mass. Motivated by the results of these previous works, an approach
similar to the galaxy identification with subhalos is adopted in
this study. Despite some of our simulations include gas particles,
we shall ignore the gas component in the identification of galaxies.
Our approach in fact combines a subhalo finding algorithm, HiFOF,
and a galaxy formation model. It 'directly' locates where the galaxy
should be formed and shows agreements with various observation
results, such as galaxy mass function, two-point correlation
function, etc. This result indicates that our galaxy identification
model not only can successfully locate subhalos in simulations that
reside inside host halos in real universe but also can give a galaxy
mass function close to the observational one.

The paper is structured as follows. In section 2, we describe the
simulation and the halo-finding algorithm in use. Thereafter, the
galaxy halo model we adopt is discussed. In Section 3, we compare
our galaxy samples with observations including the differential mass
functions, the number density evolution, the correlation functions
at different redshift, the halo occupation distribution, and the
pair fraction. Finally, we discuss and summarize our results in Section
4.

\section{Theoretical models}
\subsection{Simulations}
Our simulations have been evolved in the concordance flat
$\Lambda$CDM model: $\Omega_{0}=0.3, \Omega_{\Lambda}$ = 0.7,
$\Omega_{b} = 0.05$, and $h = 0.66$, where $\Omega_{0}$,
$\Omega_{\Lambda}$, $\Omega_{b}$ are the present-day matter, vacuum,
and baryon densities and h is the dimensionless Hubble constant
defined as $H_{0}$ $\equiv$ 100 $~ h ~ km ~ s^{-1} ~ Mpc^{-1}$.
These parameters are consistent with current observations on
cosmological parameters \citep[]{spe03,teg04,sel05,san06} with
$\Omega_{dm}$ = 0.20 $\pm$ 0.020, $\Omega_{b}$ = 0.042 $\pm$ 0.002,
$\Omega_{\Lambda}$ = 0.76 $\pm$ 0.020, h = 0.74 $\pm$ 0.02. Four
realizations with different box sizes and particle numbers were
analyzed and compared to investigate the boundary and resolution
effects as well as to have better statistics. GADGET1 \citep{spri01}
and GADGET2 \citep{sprin05} were both employed to conduct the
simulations. Our simulations all started at redshift $z = 100$ and
evolved to $z = 0$, and their density spectra were normalized to
$\sigma_{8}=0.94$, where $\sigma_{8}$ is the rms fluctuation in
sphere of $8\Mpc$ comoving radius. The first simulation
($\Lambda$$CDM_{100a}$) follows the evolution of $256^{3}$ dark
matter particles and $256^{3}$ gas particles in a $100\Mpc$ box on a
side. The mass of a dark matter particle is $m_{dm}=$ 4.125
$\times10^{9} \Msun$, while the mass of a gas particle is $m_{gas}=$
8.25 $\times10^{8}$ $\Msun$. In addition, We adopted a softening
length switching from a comoving scale to a physical scale at $z =
2.3$. The softening length $\epsilon$ was 20 $\kpc$ (comoving)
before redshift $z = 2.3$. After that, $\epsilon$ was switched to
$6\kpc$ (physical). Thus, the highest force resolution is $6\kpc$.
The second simulation ($\Lambda$$CDM_{200}$) follows the evolution
of $512^{3}$ dark matter particles and $512^{3}$ gas particles in
the same cosmology but in a $200\Mpc$ box on a  side. Both dark
matter particles and gas particles have the same masses respectively
as in the first simulation, but in this run $\epsilon$ keeps
constant at $10\kpc$
 (comoving). The third simulation ($\Lambda$$CDM_{100b}$) evolves
$512^{3}$ pure dark matter particles in a 100 $\Mpc$ box on a side.
The mass of a dark matter particle is $m_{dm}=$ 6.188 $\times10^{8}
\Msun$. We also adopt a softening length switching scheme in this
simulation. However, the softening length $\epsilon$ was set to be
$10\kpc$ (comoving) before redshift $z = 2.3$ and was modified to
$3\kpc$ (physical) thereafter. The final simulation
($\Lambda$$CDM_{100c}$) was run with $512^{3}$ dark matter particles
and $512^{3}$ gas particles in a $100\Mpc$ box on a side, and
$m_{dm}=$ 5.156 $\times10^{8} \Msun$, $m_{gas}=$ 1.031 $\times10^{8}
\Msun$, and $\epsilon$ was kept constant at $10\kpc$ (comoving). The
parameters of our simulations are summarized in
Table~\ref{tab:para}.


\subsection{Hierarchical Friends-of-Friends Algorithm (HiFOF)}
There are many widely used algorithms for identifying the
substructures, such as Bound Density Maxima \citep[BDM ;][]{kly99},
SKID \citep{sta97}, and subfind \citep{spr01}. These methods were
developed to overcome the problem of identification of dark matter
halos in the very high density environments in groups and clusters.
We use a variant of Hierarchical Friends-of-Friends Algorithm
\citep[HFOF ;][]{kly99} for the halo substructure identification. To
distinguish from HFOF, we call our method HiFOF. The main difference
between the HiFOF and the HFOF is the way to determine a stable
subhalo. The HFOF uses a particle distribution at an earlier epoch
and checks both the existence and one-to-one correspondence of the
progenitor particle clusters. A candidate subhalo is considered to
be stable if it has one (or two) progenitor(s) and if it is the only
descendant of the progenitor(s). The HiFOF examines the virial
condition of subhalos and identifies the virialized subhalos at the
highest level as the stable ones. The detail is given below.

The Friends-of-Friends \citep[FOF; see, e.g.,][]{dav85} algorithm as
a base of HiFoF identifies virialized cluster halos by using a
linking length of 0.2 $ \overline{l}$ to link particles as a group
if the separation of two particles are shorter than the linking
length; here, $\overline{l} = n_{0}^{-1/3}$ and $n_{0}$ is the mean
particle density in the simulation box. However, FOF is not capable
of finding substructures in cluster halos. Our algorithm, HiFOF,
applies the FOF algorithm with a hierarchical set of linking lengths
plus a virial condition check on all identified groups on all level.
The detail description for HiFOF is as follows. We construct sixteen
hierarchical levels by decreasing progressively from the linking
length 0.2 $ \overline{l}$, the lowest level, to 0.04
$\overline{l}$, the highest level. Because of different mass
resolution in our simulations, we setup up two different criterions
for the minimum particle number allowed for a group. In the $\Lambda
CDM_{100a}$ and the $\Lambda CDM_{200}$, the minimum particle number
for a group searched is set to be ten, $n_{min}\geq 10$,
corresponding $4.1\times 10^{10} \Msun$. As in the $\Lambda$$CDM_{100b}$ and the $\Lambda$$CDM_{100c}$, $n_{min}$ is set to be greater than 30, corresponding to $1.9\times 10^{10} \Msun$ and $1.55\times 10^{10} \Msun$, respectively. To examine the resolution effect due to $n_{min}$, the analysis in $\Lambda$$CDM_{100c}$ with $n_{min} = 10$, corresponding to $5.15 \times 10^{9} \Msun$, is additionally included and marked $^{*}\Lambda$$CDM_{100c}$. During the hierarchical tree construction, any group with particle number less than $n_{min}$ is discarded. For
every cluster halo, once a tree of sixteen levels are built, a
virial condition check is then performed on every member in the
tree. The virial condition is calculated by summing up the potential
energy and kinetic energy in physical coordinates of all particles
in a group. A parameter for the virial condition check is defined as
\begin{equation}\label{beta}
 \beta =2 \frac{\displaystyle\sum_{i} E_{kinetic}(i)}{\displaystyle\sum_{i} E_{potential}(i)} + 1
\end{equation}
where $i$ sums over the entire particles in a subhalo. It is known
that for a virialized system the virial parameter $\beta$ is equal
to 0. However, the virial condition for subhalos within a host halo
could be made less restrictive due to the potential well of the
subhalo residing inside an even deeper host halo potential.
Nevertheless, it is difficult and impractical to really calculate
the extra potential energy provided by the host halo. We therefore
adopt a loose constraint that $\beta$ has to be greater than -3;
that is, we allow a particle group with $E_{kinetic} \le 2
|E_{potential}|$ to pass the virial condition. The samples adopting
this loose constraint turn out to do better in the comparisons with
observations than the ones adopting the original condition, $\beta =
0$. For example, in the $\Lambda CDM_{100a}$ at $z = 0.3$ with
minimum galaxy mass $4.125 \times 10^{10} \Msun$, the identified
galaxy (see detail in Section~\ref{gmodel}) number density for $\beta =
-3$ is $2.01\times 10^{-2} \den$ and for $\beta = 0$ is $0.26 \times
10^{-2} \den$. It can be seen that the galaxy density with $\beta =
0$ is much lower than the one with $\beta = -3$, and a similar
situation appears in the galaxy mass functions.

\begin{figure*}
 \begin{center}
  \includegraphics[width=15cm]{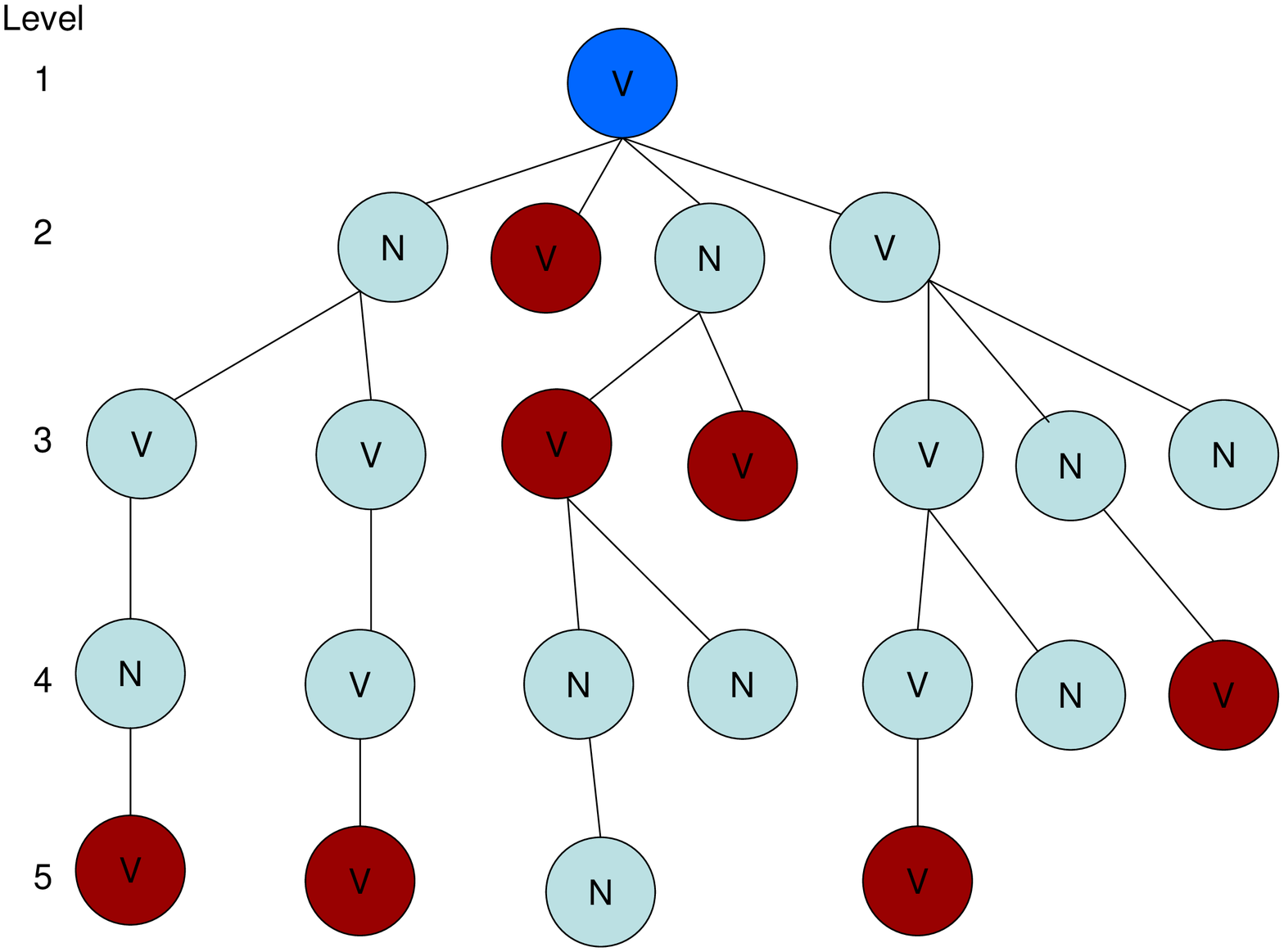}\\
  \caption{Example of the HiFOF algorithm is illustrated with a five-level hierarchical tree structure of one FOF halo.
            Points with "V" stand for "virialized" halos whereas points with "N" are for
            "unvirialized" halos. Blue, brown, and light-blue points represent
            the FOF halo, the selected galactic subhalos, and the subhalos dropped. As mentioned in the text,
            the hierarchical tree is constructed through the FOF method
            with gradually reduced linking lengths. To determine whether a subhalo is bound,
            the virial condition of it is checked. We start with selecting the
            samples from virialized subhalos on the highest level; once a
            member in a branch is selected, others in that branch will be
            discarded. We then move to the next lower level and repeat the
            procedures just described. A completed subhalo sample is then acquired after every level is examined.
   }\label{fig:HiFOF}
 \end{center}
\end{figure*}

Additionally, unlike the unbound procedure that iteratively removes
the unbound particles with the greatest energy until only bound
particles remain in the subhalo \citep{spr01}, our subhalos are
treated as a whole and can not be separated into bound and unbound
particles. In other words, if a subhalo can not pass the virial
condition check, we will drop it entirely. After the virial
conditions of all subhalos in the tree are checked, the virialized
subhalos on every branch at the highest possible level are then
selected to be the samples used in our analysis; other unvirialized
subhalos and virialized subhalos not at the highest possible level
are dropped. Figure~\ref{fig:HiFOF} schematically shows how the
HiFOF algorithm works. Continuing the same procedure on all FOF
halos, subhalo samples are then constructed. This full subhalo
sample is also called the HiFOF sample in this study.

\begin{figure}
  \begin{center}
   \leavevmode
  \includegraphics[width=9.0cm]{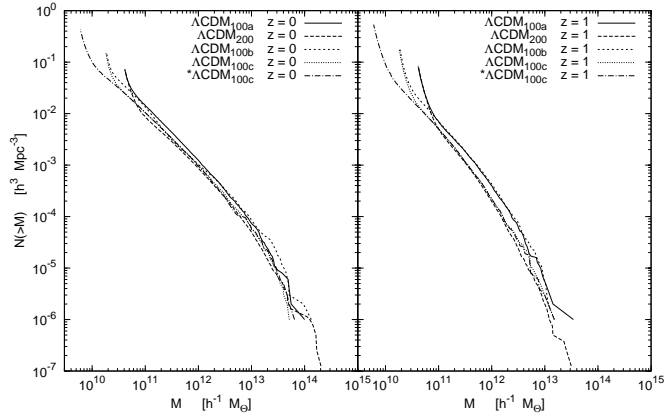}\\
  \end{center}
  \caption{ Cumulative mass functions of (bright and dark) subhalo samples obtained in our analysis at redshift z = 0 (left panel) in the $\Lambda CDM_{100a}$ (solid line), $\Lambda CDM_{200}$ (long-dashed line)
  , $\Lambda CDM_{100b}$ (short-dashed line)
  , and $\Lambda CDM_{100c}$ (dotted line) and at $z$ = 1 (right panel). The mass range plotted
  are down to the minimum mass of $4.1\times10^{10} \Msun$ in the $\Lambda CDM_{100a}$
  and $\Lambda CDM_{200}$, to that of $1.9\times10^{10} \Msun$ in the $\Lambda CDM_{100b}$,
  to that of $1.52\times10^{10} \Msun$ in the $\Lambda CDM_{100c}$, and to that of $5.16\times10^{9} \Msun$ in the $^{*}\Lambda CDM_{100c}$. These mass functions are generally consistent with
  each other. As a result, we believe that the influence of the box size and resolution effect
  are not important for our HiFOF samples.
  }
   \label{fig-AMF}
\end{figure}

Figure~\ref{fig-AMF} shows the cumulative mass functions of (bright and dark) subhalos
for the four simulations at redshifts $z = 0$ (left) at $z = 1$
(right). The densities are $7.11\times10^{-2} \den$ ($\Lambda
CDM_{100a}$), $7.33\times10^{-2} \den$ ($\Lambda CDM_{200}$),
$1.53\times10^{-1} \den$ ($\Lambda CDM_{100b}$), $1.53\times10^{-1}
\den$ ($\Lambda CDM_{100c}$), and $4.37\times10^{-1} \den$
($^{*}\Lambda CDM_{100c}$) at $z = 0$. The accumulated subhalo mass
functions in our simulations basically are consistent with each
other. In other words, at the same redshift the mass function is
mildly dependent on the box size and the mass resolution. However,
it is observed that the rapid increase appears at the low mass end
and is possibly due to the resolution effect when different
resolution runs are compared. At $z = 1$, the four simulations agree
with each other to a lesser degree. It indicates that the difference
likely arises from the different softening-length adopted as well
as from sample variance.

\subsection{Galaxy Model} \label{gmodel}
Galaxy formation involves complicated processes. Despite that,
cooling is essential to lower the specific entropy in the gas,
thereby increasing the gas density. Bremsstrahlung and line coolings
are the most efficient cooling mechanisms for this purpose
\citep[e.g., ][]{kaz96}. However, these cooling mechanisms require the gas temperature to
be above or close to the ionization temperature. When a self-gravitating gas is
cooled, it can also be heated by adiabatic contraction. It is thus
possible for the gas to maintain a temperature above the threshold
cooling temperature during the contraction, thereby yielding to
cooling runaway to form stars. In our galaxy model, we hence
demand that the host halo potential must be sufficiently deep for
the gas to get above the ionization temperature. This
condition is translated to a threshold host halo mass, only above
which stars can form within the host halo.

 Now, consider a halo with a virial mass $M$ and a virial radius $R$, $M$ can be related
to $R$ as
\begin{equation}\label{equ-MandR}
 M = \frac{4\pi}{3} \rho_{vir} R^{3} = \frac{4\pi}{3} \Delta_{c}(z) \rho_{bg} R^{3},
\end{equation}
where $\rho_{vir}$ is the virial density, $\Delta_{c}(z)$ the mean
overdensity of spherically virialized objects formed at redshift
$z$, and $\rho_{bg}$ the background density $\equiv
\rho_{0}(1+z)^{3}$. The overdensity $\Delta_{c}(z)$ at each redshift
can be evaluated using a fitting formula by \cite{kit96}. For
example, $\Delta_{c} \approx 335$ at $z = 0$ and $\Delta_{c} \approx
201$ at $z = 1$ for our cosmology, $\Omega_{0} = 0.3$ and
$\Omega_{\Lambda} = 0.7$.

In order to facilitate efficient gas cooling, a deep gravitational
potential is needed for the virial temperature to exceed roughly the
ionization temperature, i.e. $kT_{vir} \gtsim \frac{b}{2}\alpha^{2}
m_{e}c^{2}$, where $\alpha$ is the fine structure constant, $m_{e}$
the electron mass, and $c$ the speed of light, and $b$ has is a fudge factor
of order unity. The virial temperature for a star-forming halo can be obtained as
\begin{equation}\label{equ-T}
 kT_{vir} \approx \frac{GMm_{p}}{5R} \gtrsim \frac{b}{2}\alpha^{2} m_{e}c^{2}
\end{equation}
where G is the gravitational constant, k the Boltzmann constant, and
$m_{p}$ the proton mass. Using Equation~(\ref{equ-MandR}) to replace $R$
by $M$, it follows that
\begin{displaymath}
M \gtrsim M_{c} \approx (\frac{(\frac{5b}{2} (\frac{3}{4\pi})^{1/3}
\alpha^{2} m_{e} c^{2})}{(\Delta_{c}(z) \rho_{bg})^{1/3}G
 m_{p}})^{3/2}
\end{displaymath}
\begin{equation}\label{equ-mh}
  \approx 7.0 \times 10^{12} ~ (\Delta_{c}(z) (1+z)^{3})^{-1/2} \Msun,
\end{equation}
where a value of $ b \approx 0.8$ has  been adopted. At $z = 0$, the
threshold halo mass $M_{c} \approx 4.0 \times 10^{11} \Msun$. We
apply the mass threshold to select those FOF halos, which host our galaxy
samples. All other subhalos hosted by less massive FOF halos are considered dark, void of star formation.

\section{Simulation Results and Observations}
In this section we will compare our galaxy samples with the observed
galaxies. Comparisons include the differential mass function, the
galaxy number density evolution, the two point correlation function
at low and high redshifts, the halo occupation distribution, and
finally the kinetic pair fraction.

\subsection{Differential Mass Function}
\label{dmf}

Luminosity function (LF for short) $\phi(L)$ gives the relative
numbers of galaxies of different luminosities, and is so defined
that $\phi(L)dL$ is the number of galaxies in the luminosity
interval $L \rightarrow L + dL$ per unit volume of the Universe. The
luminosity function of galaxies can be fitted by Schechter's (1976)
formula,
\begin{equation}
  \label{equ:lumi_func}
   \phi(L)dL =
   \phi^{\star}(\frac{L}{L_{\star}})^{\alpha}exp(\frac{-L}{L_{\star}})\frac{dL}{L_{\star}}.
\end{equation}
We adopt $\phi^{\star} = (26.39^{+1.81 }_{-1.62}) \times 10^{-4}$
$Mpc^{-3}$, $\alpha=-1.30$, and $L_{\star}=10^{(-(M^{\star}+M_{\odot
B})/2.5)}L_{\odot}=1.987\pm0.25\times 10^{10}L_{\odot}$ for
$M^{\star} = -21.07\pm0.13$ from DEEP2 B-band galaxies at $z \sim
0.3$ \citep{fab07} for comparison later in our analysis. The mass
function and luminosity function are related by $F(M) dM =
\phi(L)dL$. If L can be expressed as a function of M when there is
no scatter between $M$ and $L$, the mass function $F(M)$ can be derived from the luminosity function $\phi(L)$. \cite{hoe05}
measured the weak-lensing signal as a function of rest-frame B-, V-,
and R-band luminosities for a sample of ``isolated'' galaxies from
the Red-Sequence Cluster Survey with photometric redshifts $0.2 < z
< 0.4$. They fit the measurements with a power-law for the
mass-to-light ratio
\begin{equation}
  \label{equ:mtol}
   M = M_{fid}(\frac{L}{10^{10} h^{-2} L_{\odot}})^{\beta},
\end{equation}
where $M_{fid}$ is the virial mass of a fiducial galaxy of
luminosity $L = 10^{10}$ $h^{-2}L_{x,\odot}$, and x indicates the
relevant filter. In B band, they obtained $M_{fid} =
9.9^{+1.5}_{-1.3}\times$ $10^{11} \Msun$ and $\beta = 1.5\pm0.3$.
With the power law form of the mass-to-light ratio, $F(M)$ can be
found as follows,

\begin{equation}
  \label{equ:mf}
  F(M) = \frac{\phi^{\star}(10^{10})^{\alpha +
1}}{L_{\star}^{\alpha+1}\beta M_{fid}}
   \times (\frac{M}{M_{fid}})^{\frac{\alpha+1}{\beta}-1}
   \times \exp^{-(\frac{M}{M_{fid}})^{1/\beta} \frac{10^{10}}{L_{\star}}} .
\end{equation}

\begin{figure}
  \begin{center}
    \leavevmode
      \includegraphics[width=10.0cm]{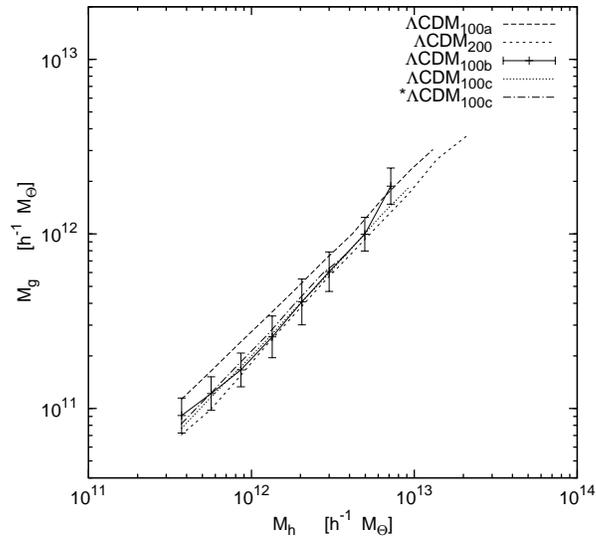}\\
    \end{center}

       \caption{Diagram shows $M_{h}$ (Mass of the isolated FOF
       halos) vs. $M_{g}$ (Mass of the HiFOF identified subhalos) at redshift $z \sim 0.3$.
       As shown in the figure, the relation can be approximated as
       a power law. This power law is used to provide the correction on galaxy mass from the weak lensing mass. The error bars represent 1 $\sigma$ values, and similar bars also exist for other curves, but omitted for the clarity of presentation.
       }
     \label{fig:mhmg}

\end{figure}

However, the galaxy mass measured by the weak lensing signal may in
fact not represent the true galaxy mass correctly. Notably, the weak lensing mass may include the mass of the host halo (FOF halo) of an isolated galaxy. We select those from our sample galaxies with only a single HiFOF subhalo residing in an FOF halo in our simulations, i.e. the isolated galaxy, to suit the observation requirement. These subhalos are a small population of the entire galaxy sample and are used as a mass calibrator. The ratios of the selected isolated galaxies to the whole galaxy sample are 32\% for $\Lambda CDM_{100a}$, 16\% for $\Lambda CDM_{100b}$, 27\% for $\Lambda CDM_{200}$, and 12\% for $\Lambda CDM_{100c}$. The result is shown in Figure~\ref{fig:mhmg} and the 1 $\sigma$ error bars are plotted only for $\Lambda CDM_{100b}$. It is found that the relation reveals a power-law form. The power law form follows
\begin{equation}
  \label{equ:mgmh}
   log_{10}(M_{g}) = a * loh_{10}(M_{h}) + b,
\end{equation}
where $a = 1.00621$ and $b = -0.806326$ for the average on all simulations. Extrapolating this power law relation to smaller galaxy mass subhalos, we apply this relation to our
entire galaxy samples.

\medskip

\begin{figure}
  \begin{center}
  \leavevmode
  \includegraphics[width=8.0cm]{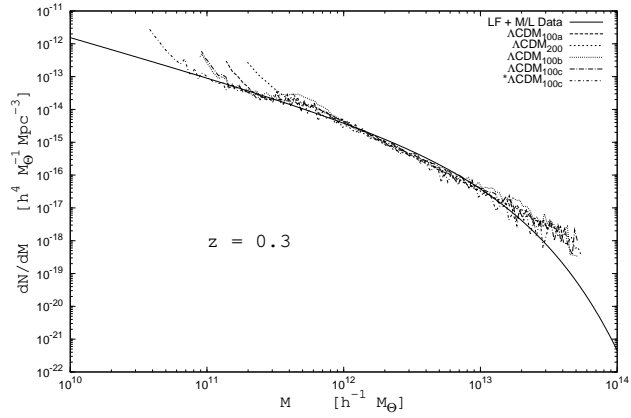}\\
  \end{center}
  \caption{Differential mass functions at redshift $z = 0.3$ of LF + M/L data (solid line),
      the $\Lambda CDM_{100a}$ samples (long-dashed
     line), $\Lambda CDM_{200}$ (short-dashed line)
     , $\Lambda CDM_{100b}$ (dotted line), $\Lambda CDM_{100c}$ dashed-dotted line)
     , and $^{*}\Lambda CDM_{100c}$ (short-dashed dotted line).
     The mass of our galaxy samples is modified according to the
     relation in Fig~\ref{fig:mhmg}. We list the space density as
     follows.
     The number density $n = 2.01\times10^{-2} \den$ in $\Lambda CDM_{100a}$,
     $2.11\times10^{-2} \den$ in $\Lambda CDM_{200}$,
     $3.22\times10^{-2} \den$ in $\Lambda CDM_{100b}$, $2.82\times10^{-2} \den$ in $\Lambda CDM_{100c}$,
     and $4.77\times10^{-2} \den$ in $^{*}\Lambda CDM_{100c}$.
     Our data profiles are similar to the observational
     data line despite deviations shown in the two ends, the lower and higher mass range.}
     \label{fig-HFOFMF}
\end{figure}

Figure~\ref{fig-HFOFMF} shows differential mass functions derived from the
luminosity function and the M/L relation in Equation~(\ref{equ:mf})
and our galaxy samples at $z = 0.3$ after applying the $M_{g}-M_{h}$
power law relation in Figure~\ref{fig:mhmg} to our galaxy samples. The mass
functions of four simulations agree with the observation, and also with each other, quite well. In highest mass range, $M > 10^{13} \Msun$, and in the low mass end, the profiles of our mass functions show the excess compared with the observational data.

In order to understand the low mass excess, a galaxy mass cut with $n_{min} = 10$ denoted as $^{*}\Lambda CDM_{100c}$, in contrast to $n_{min} = 30$ for $\Lambda CDM_{100c}$, is analyzed. We find that the low-mass excess is pushed toward lower mass. Hence, it is expected that the low-mass excess can be pushed to the very low mass end if the simulation mass resolution approaches infinitely high. As for the high-mass excess, it is likely related to the over-abundance of CD galaxies at the cluster centers, whose population deviates from the Schechter's function.

\subsection{Galaxy Density Evolution}
\label{gde}
As the high redshift data are gradually gathered in
recent years, study of galaxy properties in time evolution becomes
feasible. An aspect to test our galaxy model is to make
comparison with the observed galaxy density evolution.
When we integrate Equation~(\ref{equ:lumi_func}) over luminosity, it gives the galaxy number density,
  \begin{equation}\label{lumi_number}
    n_{(>L)} = \int^{\infty}_{L} \Phi(L^{\prime}) dL^{\prime}.
  \end{equation}
That is, with given $\phi^{\star}$, $\alpha$, $L_{\star}$, and a
luminosity cut $L$, the galaxy number density can be obtained. It is suggested by \cite{con05} that isolated
galaxies at $z \sim 1$ have a similar mass as isolated galaxies
that are 1 mag fainter at $z \sim 0$. They found that there has been
little or no evolution in the halo mass of isolated galaxies with
magnitudes in the range $\sim M_{B}^{\star}-0.5~to~-1.5$, even
though $M_{B}^{\star}$ has evolved by $\sim 1$ mag over this
redshift range. This result is adopted in our analysis. However, \cite{con05} also
assume that there is no evolution for isolated galaxies with
magnitudes $\sim M_{B}^{\star} + 2.5\log_{10}4$, which is equivalent
to $L_{\star}/4$. The luminosity cut $L$ in Equation~(\ref{lumi_number})
for $n_{(>L)}$ is set to be greater than $L_{\star}(z)/4$.
That is, the cut is $L_{\star}(z)$-dependent. In contrast, we need to find the corresponding redshift-independent
mass-cut from the simulations to make a correct comparison if the previous result and the
assumption are to be adopted. Unlike others using the galaxy density
as an indicator to provide a certain absolute magnitude threshold so as to
determine the galaxy mass, we take a different approach. We combine the LF of the DEEP2 B-band galaxies and the mass-to-light relation at $z \sim 0.3$, as well as the $M_{g}-M_{h}$ relation
discussed in Section~\ref{dmf} to obtain the corresponding mass-cut. The
luminosity cut of $L_{\star}(z \sim 0.3)/4$ given from the observed DEEP2 LF is then converted to a galaxy mass-cut in our samples with $7.02 \times 10^{10} \Msun$ in the $\Lambda
CDM_{100a}$, $7.06 \times 10^{10} \Msun$ in the $\Lambda CDM_{200}$,
$7.01 \times 10^{10} \Msun$ in the $\Lambda CDM_{100b}$ , $6.66
\times 10^{10} \Msun$ in the $\Lambda CDM_{100c}$, and $5.79 \times
10^{10} \Msun$ in the $^{*}\Lambda CDM_{100c}$. The mass cut is
applied to all redshift and the redshift evolution of the galaxy
number density can then be obtained for our galaxy samples.

\begin{figure}
  \begin{center}
    \leavevmode
      \includegraphics[width=10.0cm]{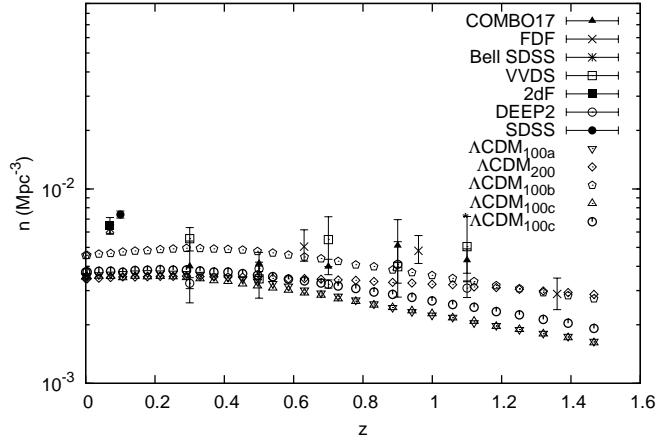}\\
   \end{center}

       \caption{ Galaxy density as a function of the
       redshift is plotted. Observational data are based on the data of the LFs in ~\cite{fab07}
       and can be converted to the galaxy density evolution by using Equation~(\ref{lumi_number}).
       A proper luminosity cut, $L \geq  L_{\star}(z)/4$, is adopted for the density calculation, and
       combining the data of the DEEP2 B-band galaxies and the mass-to-light relation at $z ~\sim 0.3$, as well as the $M_{g}-M_{h}$ relation,
       a corresponding mass-cut can be obtained and then applied to galaxy samples at other redshifts to obtain $n(z)$.
        }
     \label{fig:gden}

\end{figure}

Figure~\ref{fig:gden} presents the number density evolution of our
data and the observed galaxy density data by integrating the
luminosity functions (Equation~\ref{lumi_number}) of ~\cite{fab07}, which
include measurements of Combo-17 \citep{fab07}, FDF \citep{gab04},
Bell SDSS \citep{bel03}, VVDS \citep{ilb05}, 2df \citep{nor02},
DEEP2 \citep{fab07}, and SDSS \citep{bla03}. The profiles of our
galaxy density are basically in broad agreement with the observed
galaxies. Our data show the same evolutionary trend, a decline with
the redshift. However, at low redshift, a depletion is found in our simulations. The depletion
can probably be attributed to the failure of the assumption for
the redshift-independent mass-cut at low redshift, and can be
corrected if a different mass-cut at low redshift is assumed.

\subsection{Two-Point Correlation Function}
 The two-point correlation function
$\xi(r)$ (CF for short) is the most used indicator to quantify the
degree of clustering in a galaxy sample. It is defined as a measure
of the excess probability above Poisson for finding an objects in a
volume element $dV$ at a separation $r$ from an otherwise randomly
chosen object,

\begin{equation}\label{eq-cf-prof}
    dP = n[1+ \xi(r)]dV,
\end{equation}
where $n$ is the mean number density of the object in question. It
follows a simple power law form,

\begin{equation}\label{eq-cf}
    \xi(r) = (\frac{r}{r_{0}})^{-\gamma},
\end{equation}
where $r_{0}$ is the correlation length and $\gamma$ is the power
index of CF.

\begin{figure}
  \begin{center}
    \leavevmode
      \includegraphics[width=9.0cm]{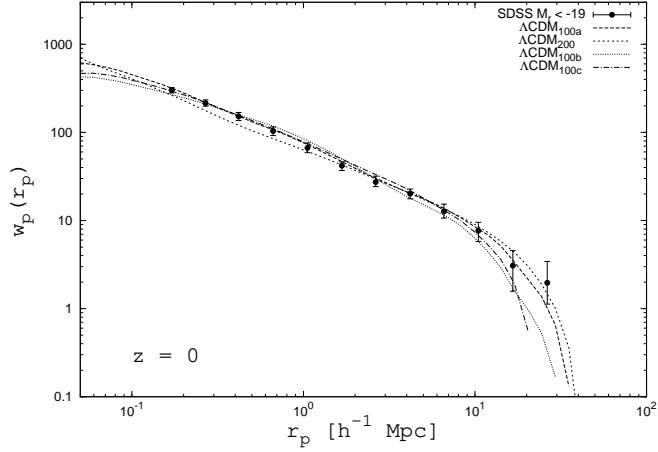}\\
   \end{center}
       \caption{Projected correlation functions of the galaxy samples at $z =
       0$ in the $\Lambda CDM_{100a}$ (dashed line), $\Lambda CDM_{200}$ (dotted line),
       $\Lambda CDM_{100b}$ (open triangle with the error bar), and $\Lambda CDM_{100c}$ (dashed-dotted line)
       are compared with the projected correlation function of the volume limited sample
       ($M_{r} < -19$) of the SDSS galaxy \citep{zeh05} with the density $1.507 \times 10^{-2}
       \den$.
       }
     \label{fig:cfatz0}
\end{figure}

\begin{figure}
  \begin{center}
    \leavevmode
      \includegraphics[width=9.0cm]{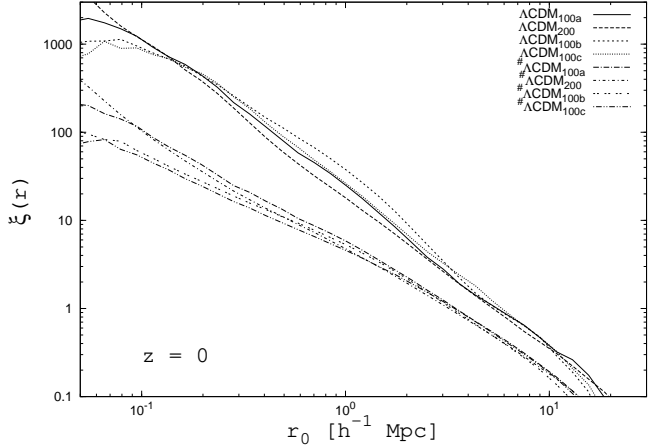}\\
   \end{center}
       \caption{The correlation functions $\xi(r)_{gg}$ of the galaxy samples at $z =
       0$ in the $\Lambda CDM_{100a}$ (solid line), $\Lambda CDM_{200}$ (dashed line),
       $\Lambda CDM_{100b}$ (short-dashed line), and $\Lambda CDM_{100c}$ (dotted line)
       are plotted. In contrast, the subhalo-subhalo projected correlation $\xi(r)_{hh}$ is presented with $\#$. It is clearly shown that $\xi(r)_{hh}$ obeys a power law form, but with a flatter slope and a smaller amplitude. That is, our galaxy samples behaves substantially differently from HiFOF samples.
       }
     \label{fig:bdcfz0}
\end{figure}

To learn about the real-space correlation function, we follow the
standard practice and compute the projected correlation function
\begin{equation}\label{equ:pcf}
    w_{p}(r_{p}) = 2 ~ \int^{r_{max}}_{0} \xi(\sqrt{r_{p}^{2} + y^{2}})
    dy.
\end{equation}
The integration limit $r_{max}$ is set to be $40 \Mpc$ for $z$ = 0
as in SDSS samples. The projected correlation functions of our
galaxy samples at $z = 0$ are shown in Figure~\ref{fig:cfatz0}. The
solid line represents the volume limited sample ($M_{r} < -19.0$) of
SDSS galaxy \citep{zeh05} with the density $1.507 \times 10^{-2}
\den$, $r_{0} = 4.56 \pm 0.23 \Mpc$, and $\gamma = 1.89 \pm 0.03$. We
select the mass cuts for our galaxy samples to match the number density of
SDSS. It is seen that both the amplitude and shape of the
projected CF at $z = 0$ are in very good agreement with those of the SDSS
data for all simulations. The close agreement of galaxy correlation functions implies
that the overall clustering of the galaxy population is determined
by the distribution of their dark matter subhalos subject to the condition of a sufficiently deep halo potential to trigger galaxy formation. In Figure~\ref{fig:bdcfz0} we plot the galaxy-galaxy CFs $\xi(r)_{gg}$ and the subhalo-subhalo CFs $\xi(r)_{hh}$ without imposing the condition of a sufficiently deep host halo potential for comparison. It clearly shows that $\xi(r)_{hh}$ also obeys a power law form, but with a flatter slope and a smaller amplitude. Note that the number density of the subhalo sample that gives $\xi(r)_{hh}$ is roughly 5 times higher than that of our galaxy subhalos for all simulations.

\begin{figure}
  \begin{center}
    \leavevmode
       \includegraphics[width=9.0cm]{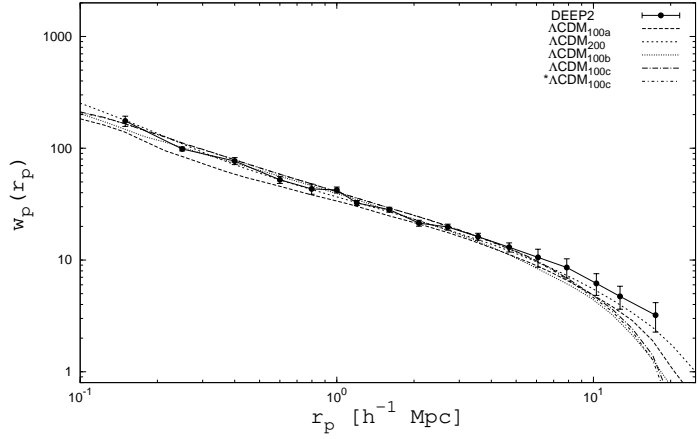}\\
      \end{center}

       \caption{ Comparison of the projected correlation functions between our four simulations at $z =1$ and the DEEP2 volume limited data of bright, $M_{B} < -19.0$,  galaxies
       \citep{coi06}. The CF of the DEEP2 data has a density $0.013 \den$.
       We therefore select our samples to have the same density, and the results can be found to
       be in good agreement with the DEEP2 data.}
     \label{fig:cfatz1}
\end{figure}

Figure~\ref{fig:cfatz1} shows the results of the projected
correlation functions at $z = 1$ of our galaxy samples for the four
simulations and the volume-limited sample of bright,
$(M_{B} \geq -19.0)$, DEEP2 galaxies \citep{coi06}. The upper
integration limit is set to $r_{max} = 20 \Mpc$ in our simulations
to agree with the DEEP2 samples. The DEEP2 galaxy sample has a
density of $1.3 \times 10^{-2} \den$. We therefore select our
samples to match the density of the DEEP2 sample. In
Figure~\ref{fig:cfatz1}, it can be seen that the correlation
functions of the DEEP2 galaxies and our samples also agree very well.

\begin{figure}
  \begin{center}
    \leavevmode
       \includegraphics[width=8.5cm]{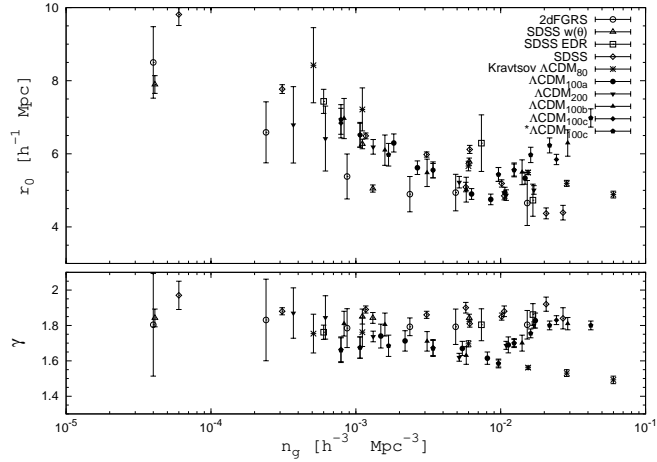}\\
  \end{center}
       \caption{Top: Best-fit correlation scale $r_{0}$ as a function of number density
             at the present-day epoch. Bottom: Best-fit slope $\gamma$ of the correlation
             function as a function of number density. This plot is based on the Figure 11 in
             \cite{kra04} plus the newest SDSS data \citep{zeh05}
             and our data. Our values of the correlation length $r_{0}$
             are consistent with observational data. The values of the power index
         $\gamma$ are lower at density around $ n = 10^{-2} \den$. However, the agreement between observations and our data is seen.
             }
     \label{fig:cfz0}

\end{figure}

\begin{figure}
  \begin{center}
    \leavevmode
      \includegraphics[width=9.0cm]{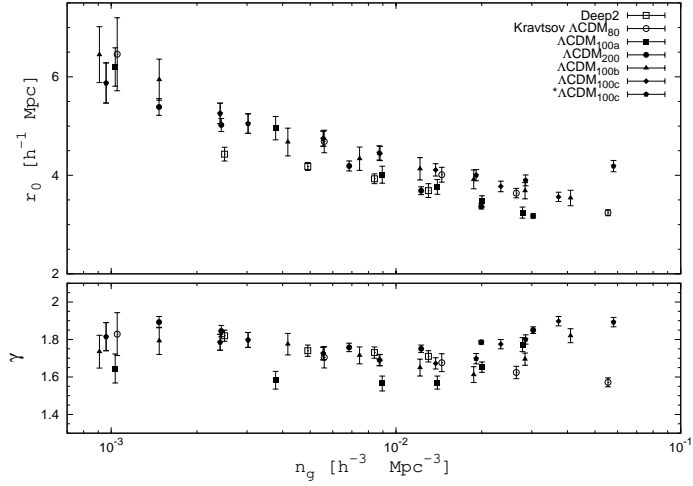}\\
  \end{center}
       \caption{Same as Figure~\ref{fig:cfz0}, but for $z = 1$.
       We additionally include DEEP2 data released by \citep{coi06} as well as our data.
       It is found that the correlation length and the power index in our simulations match DEEP2 data well.}
     \label{fig:cfz1}

\end{figure}

In Figure~\ref{fig:cfz0} and Figure~\ref{fig:cfz1}, the correlation
length $r_{0}$ and the power index $\gamma$ are plotted as a
function of different space density at $z = 0$ and $z = 1$,
respectively. The CFs are fit over the range of scales from 0.2 to
$13 \Mpc$ where the errors in $\xi(r)$ in our samples are the
"jackknife" 1 $\sigma$ errors, computed using the eight octants of
the simulation cube \citep[see][]{wei04}. Figure~\ref{fig:cfz0} is
basically adopted from Figure 11 in \cite{kra04} which includes the
Two-Degree Field \citep[2dF;][]{nor02}, SDSS galaxy surveys
\citep{zeh02,bud03} and their simulation data. We additionally add
the newest SDSS data \citep{zeh05} and our galaxy data into
Figure~\ref{fig:cfz0}. The strong dependence of correlation length
$r_{0}$ on the number density $n$ is evident in
Figure~\ref{fig:cfz0}. That is, brighter galaxies are more
clustered. In addition, a nearly constant profile of $\gamma$ in
observations is seen. The amplitude of $\gamma$ in our simulations,
in general, also presents a roughly flat trend although at $n
\approx 10^{-2} \den$ our $n$ values show slight depletion compared
with the observations. At $z = 1$ we plot the DEEP2 data released by
\cite{coi06}, the simulation results of \cite{kra04}, and our data.
A similar conclusion as at $z = 0$ is obtained for $r_{0}$, and our
values of $\gamma$ also show good consistency with the DEEP2 galaxies.

\subsection{Halo Occupation Distribution (HOD)}

 The HOD formalism, developed during the last several
years, has become a powerful theoretical framework for predicting
and interpreting galaxy clustering. The original HOD uses the
probability $P(N|M)$ to describe the bias of a class of galaxies
that a halo of virial mass M contains N such galaxies. ~\cite{ber03}
studied HOD and found that at a given halo mass it is statistically
independent of the halo¡¦s large-scale environment. In addition,
they compared HOD of a semi-analytic model and of gasdynamics
simulations in detail and concluded that the semi-analytical HOD for
samples of the same space density agree remarkably well with simulations, despite that the
two methods predict different galaxy mass functions.

~\cite{kra04} showed that HOD can actually be understood as a
combination of the probability for a halo of mass M to host a
central galaxy and the probability to host a number of satellite
galaxies that obeys Poisson statistics. We analyze the first moment
of HOD, $\langle N(M_{h}) \rangle \equiv \langle N_{c}(M_{h})
\rangle  + \langle N_{s}(M_{h}) \rangle$, as a function of host mass
$M_{h}$ for the halo samples, where $\langle N_{c} \rangle$ is the
HOD of central galaxies modeled as a step function $\langle N_{c}
\rangle = 1$ for $M > M_{min}$ and $\langle N_{c} \rangle = 0$
otherwise, and $\langle N_{s} \rangle$ is number of the satellite
galaxies modeled as a power law $ \langle N_{s} \rangle \propto
M^{\alpha}$ suggested by \cite{kra04}.  \cite{zeh05} used this HOD framework to interpret their results. Due to a finite (but important) subset of information encoded in the correlation function
for galaxy clustering, a restricted HOD model was employed and fitted with a
small number of free parameters. The HOD formulation they implemented has three free
parameters: $M_{min}$, the minimum halo mass for galaxies above the
luminosity threshold, $M_{1}$, the mass of a halo hosting one
satellite galaxy above the luminosity threshold, and $\alpha$, the
power law slope of the satellite mean occupation function. They took
$M_{min}$ to be fixed by matching the observed space density of the
sample, leaving $M_{1}$ and $\alpha$ as free parameters to fit their
projected CFs. More recently, \cite{zhe07} modeled the
luminosity-dependent projected two-point correlation function of
DEEP2 $(z \sim 1)$ and SDSS $(z \sim 0)$ galaxies within the HOD
framework. They adopted a more flexible parameterizations with five
parameters, motivated by the less satisfactory results of
\cite{zhe05} as well as of the three-parameter parameterizations.
In the following test on this aspect, only the three-parameter parametrization scheme is adopted for simplicity.

\begin{figure}
  \begin{center}
    \leavevmode
      \includegraphics[width=8.5cm]{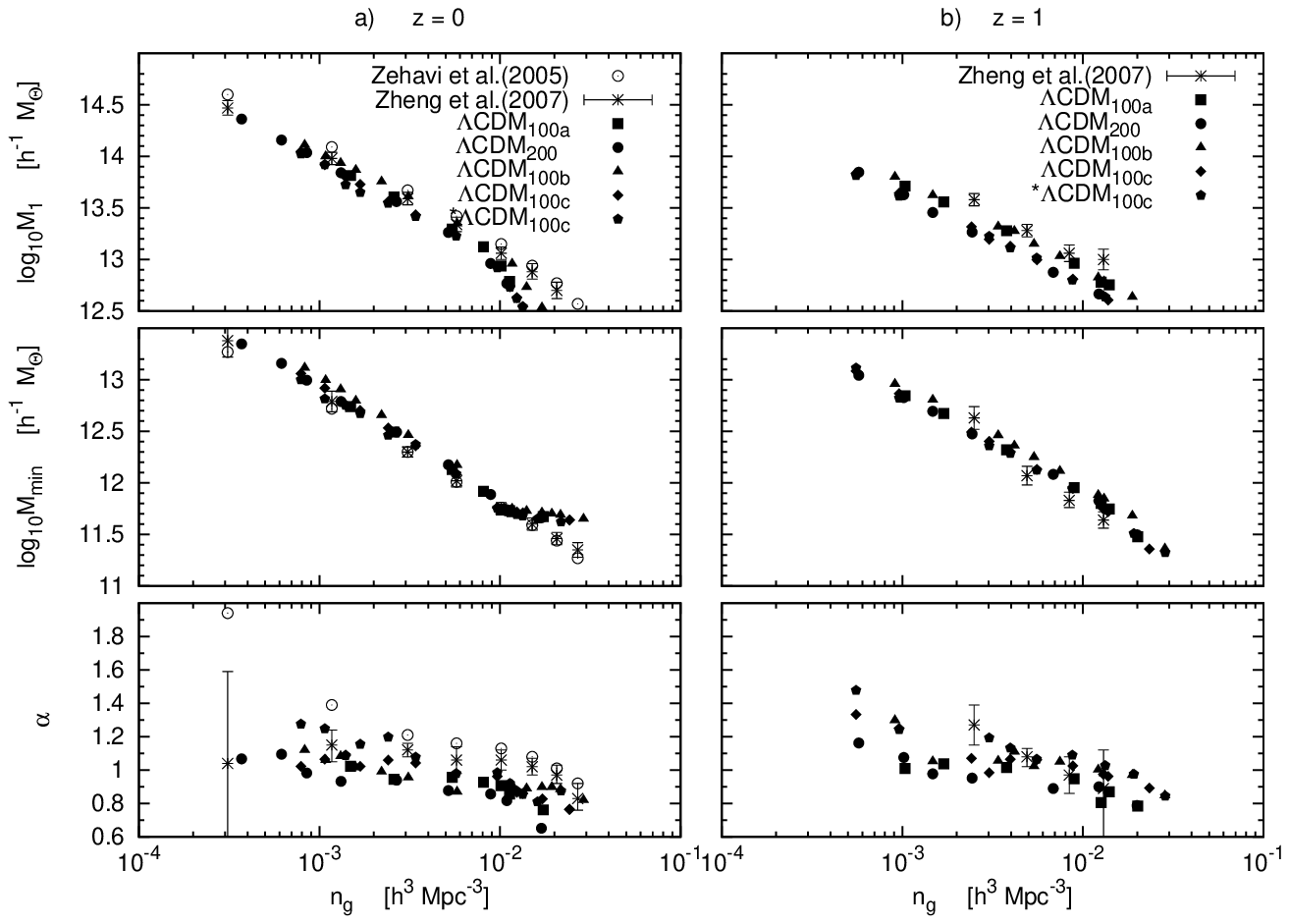}\\
  \end{center}
       \caption
       {From top to bottom, the three fitting parameters $M_{1}$, $M_{min}$,
        and $\alpha$ of the HOD formulation are plotted as a function
        of the galaxy number density $n_{g}$. In \ref{fig:sdsshod}(a), the open circles  represent
        SDSS data \citep{zeh05} and the stars with errorbars give also SDSS data \citep{zhe07},
        and in \ref{fig:sdsshod}(b) the stars with errorbars give DEEP2 data \citep{zhe07},
        while the filled square, circle, triangle,
        diamond, and pentagon are for the $\Lambda CDM_{100a}$, $\Lambda CDM_{200}$,
        $\Lambda CDM_{100b}$, $\Lambda CDM_{100c}$, and $^{*}\Lambda CDM_{100c}$ data.
        Consistency between SDSS galaxies and our galaxy samples is observed at
        $z = 0$. The good agreements are also seen between DEEP2 galaxies and ours at $z = 1$.}
     \label{fig:sdsshod}

\end{figure}

We compute the HOD of our galaxy samples to obtain $M_{min}$,
$M_{1}$, and $\alpha$ as a function of galaxy number density to make
comparisons with SDSS and DEEP2 galaxies. In
Figure~\ref{fig:sdsshod}$(a)$, from top to bottom, our results of
$M_{min}$, $M_{1}$, and $\alpha$ in the four simulations at $z = 0$
and those of SDSS galaxies are compared. It can be seen that
$M_{min}$ and $M_{1}$ of four simulations not only agree with each
other, implying the lesser degree of dependence on the box size and
the resolution, but also match the observation quite well no matter
in the magnitude or in the trend. Despite the increasing trend of the
slope $\alpha$ with a decreasing density is similar to that of SDSS
galaxies, the $\alpha$ value of our HODs on the whole appears to be
slightly smaller than that of the observations. This may results
in slightly insufficient subhalos found in massive clusters in our samples.
Figure~\ref{fig:sdsshod}$(b)$ shows our results at $z = 1$ and the
observational data of DEEP2 galaxies. As on can see, the agreements
on $M_{min}$, $M_{1}$, and $\alpha$ are also good.

\subsection{Pair Fraction}
Since our galaxy sample can well reproduce the observed evolution of
galaxy number density, the mass function, the 2-point correlation function and the HOD, we therefore make an attempt to put another comparison to the simulation resolution limit by exploring the pair fraction and merger rate.

DEEP2 Team ~\citep{lin04} explored the kinematic close pair fraction
and the merger rate up to redshift $z \sim 1.2$ and disclosed weak
evolution in the galaxy pair fraction. Assuming mild luminosity
evolution, they found the number of companions per luminous galaxy
to evolve as $(1 + z)^{m}$, with m = 0.51 $\pm$ 0.28 for the
$r_{max} = 50 \kpc$ case. Recently, ~\cite{lin08} used more complete
data to study the same problem, and obtained an improved result
with m = 0.41 $\pm$ 0.20 for all galaxies. These two studies
consistently reveal that the pair fraction of galaxies indeed
undergo weak evolution.

We study the pair fraction problem with our
galaxy samples; our approach is in variance with applying a
"hybrid" formalism to address this problem \citep{ber06}, which
combines an N-body simulation to account for large-scale structure
and the host dark matter halo population and an analytic
substructure model \citep{zen05} to identify satellite galaxies
within the host halos. We test our samples with the same setup as in \cite{lin04} and
\cite{lin08}. They defined close pairs such that the projected
separations satisfy $10 \kpc$ $\leq$ $\Delta r$ $\leq$ $r_{max}$,
where $r_{max}$ = 30, 50, or 100$\kpc$, and the rest-frame relative
velocity $\Delta v$ less than 500 $km ~ s^{-1}$. To ensure the selection of the same types of galaxies at
different redshifts in the presence of luminosity evolution, a
specific range in the evolution-corrected absolute magnitude
$M^{e}_{B}$, defined as $M_{B} + Qz$, was adopted in our analysis where the
evolution is parameterized as $M(z) = M(z = 0) - Qz$. \cite{lin08}
adopted $Q = 1.3$ found by \cite{fab07} and restricted their analysis
to galaxies with luminosities $-21 \leq M^{e}_{B} \leq -19$ for $z$
= 0.45-1.2. Applying the mass-to-light ratio in Equation~(\ref{equ:mtol})
for B-band with $M_{fid} = 9.9^{+1.5}_{-1.3}\times$ $10^{11} \Msun$
and $\beta = 1.5\pm0.3$, we are able to convert the mass to the
luminosity. The luminosity range, $-21.39$ $\leq M_{B} \leq$
$-19.39$, at $z = 0.3$ is considered because the mass-to-light ratio
data of \cite{hoe05} was collected at an average redshift $z \sim
0.3$. The mass range at $z = 0.3$ after conversion is found to be
$2.38 \times10^{11} \Msun$ $\leq M \leq$ $3.78\times10^{12} \Msun$.
Moreover, the average number of companions per galaxy is defined as
\begin{equation}\label{euq-pf}
    N_{c} \equiv \frac{2N_{p}}{N_{g}},
\end{equation}
where $N_{p}$ is the number density of individual paired galaxies
and $N_{g}$ the number density of galaxies as suggested in \cite{ber06}. Following \cite{ber06}, we also adopt a
fiducial relative line-of-sight velocity difference of $|\Delta v|=
500$ $km ~ s^{-1}$. A close-pair cylinder volume is then constructed
by a constant $\Delta v$ cut. In our analysis, only the $r_{max}$ =
50, and $100 \kpc$ cases are considered due to the limitation of our
mass resolution.

\begin{figure}
  \begin{center}
    \leavevmode
      \includegraphics[width=9.0cm]{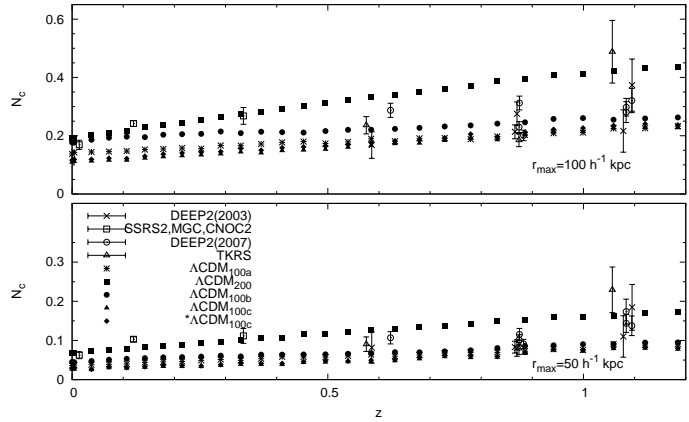}\\
  \end{center}
       \caption{Pair fraction, $N_{c}$, as a function of redshift z.
        The crosses mark the results from DEEP2 early data (2003). The open squares represents
        measurements from SSRS2, MGC, and CNOC2. The open circles show DEEP2 data (2007).
        The open triangles are from TKRS. The stars, filled squares, circles, triangles, and
        diamonds give our simulation results from $\Lambda CDM_{100a}$, $\Lambda CDM_{200}$,
        $\Lambda CDM_{100b}$, $\Lambda CDM_{100c}$, and $^{*}\Lambda CDM_{100c}$.
        Two cases of $r_{max}$, 50 (bottom) and 100 (top) $\kpc$, are considered.}
     \label{fig:pf}

\end{figure}

Figure~\ref{fig:pf} shows the redshift $z$ versus the pair fraction
for $r_{max}$ = 50 (bottom) and  $r_{max}$ = $100 \kpc$ (top). The
pair fraction data are taken from \cite{lin04} including SSRS2
\citep{dac98}, CNOC2 \citep{yee00}, and DEEP2 early data
\citep{dav03} and from full samples in \cite{lin08} including SRSS2,
CNOC2, MGC \citep[Millennium Galaxy Catalog, ][]{lis03,dri05,all06},
TKRS \citep{wir04}, and DEEP2 \citep{dav03,dav07}. Our data points
are consistent with the DEEP2 medium to high redshift observations
although at low redshift we have lower values than other
observations. However, the mild evolution trend is seen. Using the
fitting formula, $N_{c}(0)(1+z)^{m}$, proposed in \cite{lin04} and
\cite{lin08}, we find the following results. For $r_{max} = 50 \kpc$
, $N_{c}(0) = 0.036$ and $m = 1.17$ in the $\Lambda$$CDM_{100a}$ ,
$N_{c}(0) = 0.074$ and $m = 1.07$ in the $\Lambda$$CDM_{200}$,
$N_{c}(0) = 0.046$ and $m = 0.96$ in the $\Lambda$$CDM_{100b}$,
$N_{c}(0) = 0.025$ and $m = 1.43$ in the $\Lambda$$CDM_{100c}$, and
$N_{c}(0) = 0.028$ and $m = 1.37$ in the $^{*}\Lambda$$CDM_{100c}$.
For $r_{max} = 100 \kpc$, $N_{c}(0) = 0.137$ and $m = 0.65$ in the
$\Lambda$$CDM_{100a}$, $N_{c}(0) = 0.207$ and $m = 0.95$ in the
$\Lambda$$CDM_{200}$, $N_{c}(0) = 0.182$ and $m = 0.48$ in the
$\Lambda$$CDM_{100b}$, $N_{c}(0) = 0.107$ and $m = 0.98$ in the
$\Lambda$$CDM_{100c}$, and $N_{c}(0) = 0.116$ and $m = 0.92$ in the
$^{*}\Lambda$$CDM_{100c}$. Apparently, our pair fractions show
stronger evolution than those in \cite{lin04} and \cite{lin08},
where $m = 0.51 \pm 0.28$ and $m = 0.41 \pm 0.14$ for $r_{max}$ =
$50 \kpc$, and $m = 0.47 \pm 0.18$ and $m = 0.29 \pm 0.05$ for
$r_{max}$ = $100 \kpc$, but appear to agree with the results of
\cite{ber06} for $r_{max}$ = $50 \kpc$, where $m = 0.42 \pm 0.17 $
for $V_{now}$, the maximum circular velocity of the subhalo at
the current epoch, and $m = 0.99 \pm 0.14$ for $V_{in}$, the maximum
circular velocity of the subhalo when it was first accreted
into the host halo.

The steeper slope in our sample is rooted on the too-low galaxy density at low $z$, which can basically be attributed to the failure of our assumption on the redshift-independent mass-cut. We note that the pair fraction critically depends on the galaxy density and the density is determined by the luminosity cut.
Consequently, if a wrong luminosity evolution assumption is adopted, this can lead
to a high galaxy mass-cut, a lower galaxy density, and finally a lower pair fraction.
Our data at low redshift reveal this effect and result in a steeper slope.

\section{Conclusion and Discussion}

In this study we have used the HiFOF method to locate and identify
subhalos, and we advance the idea that galaxies form only in a sufficiently
massive halo due to its sufficiently high gas cooling efficiency. It is
estimated that the threshold halo mass $M_{c} \approx 4.0 \times
10^{11}$ $h^{-1} M_{\odot}$ at $z = 0$, and our galaxy samples are
constructed from the HiFOF subhalos embedded in the halos satisfying
this mass threshold. There are far more subhalos that are hosted
by halos not meeting the mass threshold. They are dark halos where
star formation cannot occur. To test our model, we have analyzed the
differential mass function, two-point correlation function (CF), the
space density evolution, the HOD, and the pair fraction from our
galaxy samples. The results are summarized as follows.

(1) Combing the luminosity function (LF) and the mass to light
relation (M/L), a differential mass function can be derived. Based
on the lensing signal from Red-Sequence Cluster Survey to determine
the galaxy mass, we have corrected the lensing mass in our galaxy
samples so as to make fair comparison. We find that the mass
functions of our four simulations agree with the LF and M/L data
very well.

(2) We compare our galaxy density evolution with the evolution of
the observational galaxy number density through the integration of a
LF with a proper luminosity cut. Our simulations basically show fair
consistency with observations, except at low-redshift. The
insufficient density is likely due to the failure of the assumption
at low redshift about the redshift-independent mass-cut. The
depletion in the galaxy number density also reflects a similar
result in pair fractions found at low-redshift.

(3) We have analyzed CFs of our galaxy subhalos at $z = 0$ and at $z
= 1$. At $z = 0$, we select our galaxies to have the density $1.507
\times 10^{-2} \den$ to make comparison with the volume limited
sample ($M_{r} < -19$) of SDSS galaxies. Both the slope and the
amplitude are consistent with the SDSS. At $z = 1$, we select
samples at $n = 1.3 \times 10^{-2} \den$ to setup the same density
condition as in the volume-limited sample of bright (corrected),
$M_{B} > -19.0$, DEEP2 galaxies. The projected correlation functions
are compared, and our profiles are similar to the one found in DEEP2
galaxies.

(4) We also study $r_{0}$ and $\gamma$ as a function of the space
density $n$. Our galaxies reveal the increase of $r_{0}$ as $n$
decreases, a similar result also found in the observation data. In
other words, the correlation length $r_{0}$ has luminosity
dependence, and the brighter the galaxies, the larger the
correlation length. On the other hand, the power index $\gamma$ of
observation data displays a nearly constant behavior over a wide
range of $n$ at $z = 0$. Despite a smaller values of $\gamma$ are
obtained around $n \approx 10^{-2}$ in our simulations, a roughly
constant trend on the whole is seen. The same results are reached at
$z = 1$ as well.

(5) We have parameterized our HODs with three parameters $M_{1}$,
$M_{min}$, and $\alpha$ as a function of the galaxy number density.
Our values of $M_{1}$ and $M_{min}$ and their increasing trend as
the galaxy density decreases show good agreement with SDSS data
\citep[]{zeh05,zhe07} and DEEP2 data \citep{zhe07}, and the
depletion of the slope $\alpha$ at $z = 0$ may be attributed to
insufficient subhalos found in massive clusters.

(6) We have analyzed the kinematic pair fraction of our galaxy
samples by selecting the mass in between $2.28 \times10^{11} \Msun$
and $3.61\times10^{12} \Msun$ which are obtained by applying the
observed mass-to-light ratio to convert the luminosity range
$-21.39$ $\leq M_{B}^{e} \leq$ $-19.39$ at $z = 0.3$. Parameterizing
the evolution of the pair fraction as $(1 + z)^{m}$, \cite{lin08}
find that when $r_{max} = 50$ $\kpc$, $m = 0.41\pm0.14$ for
the full sample. We obtain $m = 1.17$ in $\Lambda CDM_{100a}$, $m =
1.07$ in $\Lambda CDM_{200}$, $m = 0.96$ in $\Lambda CDM_{100b}$, $m
= 1.43$ in $\Lambda CDM_{100c}$, and $m = 1.37$ in $^{*}\Lambda
CDM_{100c}$. When $r_{max} = 100 \kpc$, in contrast to $0.29\pm0.05$
found by \cite{lin08}, we find $m = 0.65$ in $\Lambda CDM_{100a}$,
$m = 0.95$ in $\Lambda CDM_{200}$, $m = 0.48$ in $\Lambda
CDM_{100b}$, $m = 0.98$ in $\Lambda CDM_{100c}$ and $m = 0.92$ in
$^{*}\Lambda CDM_{100c}$. The steeper slope basically results from insufficient close pairs found at low redshift, which is interpreted as the deficient low-$z$ galaxy density
in our samples due to the failure of the assumption on the
redshift-independent mass-cut. However, the situation is less severe
for the $r_{max} = 100 \kpc$ case where milder evolution trend is more evident.

Most of the above tests for our galaxy samples provide supporting
evidence for our galaxy model.  However, this model has some
problems of its own. First, this model does not provide information of the galaxy
morphology. Second, unlike the Millennium simulation \citep{spr05},
this model does not contain any empirical ingredient to trace
individual galaxy evolution.

The solution to the first problem requires the gas component to be included in the simulation, which can reveal
the amount of residual gas angular momentum in the subhalo thereby
providing additional information of morphological types. The
complexity of the second problem is intractable from the
first-principle calculations and it needs an empirical model, much
as what was added to the Millennium, to approximate the physics
involved.

      We would like to thank Lihwai Lin and Tak-Pong Woo for many helpful
discussions and suggestions. This project is supported in part by the grant: NSC-92-2628-M-002-008-MY3 (1/3).




\clearpage

\begin{table}
  \begin{center}
  \leavevmode
  \caption{Simulation Parameters}
  \label{tab:para}
    \begin{tabular}{cccccc} \hline\hline
                       & $L_{Box}$    & $N_{dm}$   & $M_{dm}$             & $\epsilon_{z=0}$& $\sigma_{8}$ \\
  Name                 & ($h^{-1}~Mpc)$&           & ($h^{-1}~M_{\odot})$  & ($\kpc$)   &              \\ \hline
  $\Lambda$$CDM_{100a}$& 100          & $256^{3}$ & 4.125 $\times10^{9}$ & 6.0             & 0.94         \\
  $\Lambda$$CDM_{200} $& 200          & $512^{3}$ & 4.125 $\times10^{9}$ & 10.0            & 0.94         \\
  $\Lambda$$CDM_{100b}$& 100          & $512^{3}$ & 6.188 $\times10^{8}$ & 3.0             & 0.94         \\
  $\Lambda$$CDM_{100c}$& 100          & $512^{3}$ & 5.156 $\times10^{8}$ & 10.0            & 0.94         \\ \hline
    \end{tabular}
  \end{center}
\end{table}


\begin{thebibliography}{}


\bibitem[Allen et al.(2006)]{all06} Allen, P. D., Driver, S. P., Graham, A. W., Cameron, E., Liske, J. \& de Propris, R. 2006, \mnras, {\bf{371}},
2

\bibitem[Bell et al.(2003)]{bel03}Bell, E. F., McIntosh, D. H., Katz, N. \& Weinberg, M. D. 2003, \apjs, {\bf{149}},
289

\bibitem[Berlind et al.(2003)]{ber03}Berlind, A. A. et al. 2003, \apj, {\bf{593}}, 1

\bibitem[Berrier et al.(2006)]{ber06} Berrier, J. C., Bullock, J. S., Barton, E. J., Guenther, H. D.,
Zentner, A. R., \& Wechsler, R. H. 2006, \apj, {\bf{652}}, 56

\bibitem[Blanton et al.(2003)]{bla03} Blanton, M. R. et al. 2003, \apj, {\bf{592}}, 819

\bibitem[Budavari et al.(2003)]{bud03} Budavari, T. et al. 2003, \apj, {\bf{595}}, 59


\bibitem[Coil et al.(2006)]{coi06} Coil, A. L., Newman, J. A., Cooper, M. C., Davis, M., Faber, S. M.,
Koo, D. C., \& Willmer, C. N. 2006, \apj, {\bf{644}}, 671

\bibitem[Conroy et al.(2005)]{con05} Conroy, C. et al. 2005, \apj, {\bf{635}},
982

\bibitem[Conroy et al.(2006)]{con06} Conroy, C., Wechsler, R., Kravtsov, A.  2006, \apj, {\bf{647}},
201

\bibitem[da Costa et al.(1998)]{dac98} da Costa, L. N. et al. 1998, \aj, {\bf{116}}, 1

\bibitem[Davis et al.(1985)]{dav85}Davis, M., Efstathiou, G., S., Frenk, C. S., \& White, S. D. M. 1985, \apj,
{\bf{292}}, 371

\bibitem[Davis et al.(2003)]{dav03}Davis, M. et al. 2003, SPIE, {\bf{4834}}, 161

\bibitem[Davis et al.(2007)]{dav07}Davis, M. et al. 2007, \apj, {\bf{660}}, L1

\bibitem[Driver et al.(2005)]{dri05}Driver, S. P., Liske, J., Cross, N. J. G., De Propris, R., \& Allen,
P. D. 2005, \mnras, {\bf{360}}, 81

\bibitem[Faber et al.(2007)]{fab07}Faber, S. M. et al. 2007, \apj, {\bf{665}}, 265

\bibitem[Gabasch et al.(2004)]{gab04}Gabasch, A. et al. 2004, \aap, {\bf{412}}, 41

\bibitem[Hoekstra et al.(2005)]{hoe05}Hoekstra, H., Hsieh, B. C., Yee, H. K. C., Lin, H., \& Gladders,
M. D. 2005, \apj, {\bf{635}}, 73

\bibitem[Ilbert et al.(2005)]{ilb05}Ilbert, O. et al. 2005, \aap, {\bf{439}}, 863

\bibitem[Kaze et al.(1996)]{kaz96}Katz, Neal, Weinberg, David H., Hernquist, Lars, Miralda-Escude, Jordi  1996, \apj, {\bf{457}}, L57

\bibitem[Kitayama \& Suto(1996)]{kit96}Kitayama, Tetsu, Suto, Yashshi 1996, \apj, {\bf{469}},
480

\bibitem[Klypin et al.(1999)]{kly99}Klypin, A., GottlAober, S., \& Kravsov, A. V. 1999, \apj, {\bf{516}}, 530

\bibitem[Kravtsov et al.(2004)]{kra04}Kravtsov, A. V., Berlind, A. A., Wechsler, R. H., Klypin, A. A.,
Gottloeber, S., Allgood, B., \& Primack, J. R. 2004, \apj,
{\bf{609}}, 35

\bibitem[Lin et al.(2004)]{lin04}Lin, L. et al. 2004, \apj, {\bf{617}}, L9

\bibitem[Lin et al.(2008)]{lin08}Lin, L. et al. 2008, \apj, {\bf{681}}, 232

\bibitem[Liske et al.(2003)]{lis03}Liske, J., Lemon, D. J., Driver, S. P., Cross, N. J. G., \& Couch,
W. J. 2003, \mnras, {\bf{344}}, 307

\bibitem[Neyrinck et al.(2004)]{ney04}Neyrinck, M. C., Hamilton, A. J. S., \& Gnedin, N. Y. 2004,
\mnras, {\bf{348}}, 1

\bibitem[Norberg et al.(2002)]{nor02}Norberg, P. et al. 2002, \mnras, {\bf{332}}, 827

\bibitem[Riess et al.(1998)]{rie98}Riess, A. G. et al. 1998, \aj, {\bf{116}}, 1009

\bibitem[Sanchez et al.(2006)]{san06}Sanchez, A. G., Baugh, C. M., Percival, W. J., Peacock, J. A.,
Padilla, N. D., Cole, S., Frenk, C. S., \& Norberg, P. 2006, \mnras,
{\bf{336}}, 189

\bibitem[Seljak et al.(2005)]{sel05}Seljak, U. et al. 2005, \prd, {\bf{71}}, 103515

\bibitem[Spergel et al.(2003)]{spe03}Spergel, D. N. et al. 2003, \apjs, {\bf{148}}, 175

\bibitem[Springel(2005)]{sprin05}Springel, V. 2005, \mnras, {\bf{364}}, 1105

\bibitem[Springel et al.(2005)]{spr05}Springel, V. et al. 2005, \nat, {\bf{435}}, 629

\bibitem[Springel et al.(2001a)]{spr01}Springel, V., White, S. D. M., Tormen, G., \& KauRmann, G. 2001a,
\mnras, {\bf{328}}, 726

\bibitem[Springel et al.(2001b)]{spri01}Springel, V., Yoshidaa, N., \& White, S. D. 2001b, New Astronomy,
{\bf{6}}, 79

\bibitem[Stadel et al.(1997)]{sta97}Stadel, J., Katz, N., Weinberg, D. H., \& Hernquist, L. 1997, SKID

\bibitem[Tegmark et al.(2004)]{teg04}Tegmark, M. et al. 2004, \apj, {\bf{606}}, 702

\bibitem[Weinberg et al.(2004)]{wei04}Weinberg, D. H., Dave, R., Katz, N., \& Hernquist, L. 2004, \apj,
{\bf{601}}, 1

\bibitem[White et al.(1993)]{whi93}White, S. D. M., Navarro, J. F., Evrard, A. E., \& Frenk, C. S.
1993, \nat, {\bf{336}}, 429

\bibitem[Wirth et al.(2004)]{wir04}Wirth, G. D. et al. 2004, \aj, {\bf{127}}, 3121

\bibitem[Yee et al.(2000)]{yee00}Yee, H. K. C. et al. 2000, \apjs, {\bf{129}}, 475

\bibitem[Zehavi et al.(2002)]{zeh02}Zehavi, I. et al. 2002, \apj, {\bf{571}}, 172

\bibitem[Zehavi et al.(2005)]{zeh05}Zehavi, I. et al. 2005, \apj, {\bf{630}}, 1

\bibitem[Zentner et al.(2005)]{zen05}Zentner, A. R., Berlind, A. A., Bullock, J. S., Kravtsov, A. V., \& Wechsler, R. H. 2005, \apj,
{\bf{624}}, 505

\bibitem[Zheng et al.(2005)]{zhe05}Zheng, Z. et al. 2005, \apj, {\bf{633}}, 791

\bibitem[Zheng et al.(2007)]{zhe07}Zheng, Z., Coil, A. L., Zehavi, I.  2007, \apj, {\bf{667}},
760

\end{thebibliography}
\end{document}